\documentclass[letterpaper]{aa}
\usepackage{natbib}
\usepackage{graphicx}
\usepackage{color}
\usepackage{ulem}

\title{Origin of the dust emission from Tycho's SNR}
\author{
Daisuke Ishihara$^{1}$,
Hidehiro Kaneda$^{1}$,
Akihiro Furuzawa$^{1}$,
Hideyo Kunieda$^{1}$,
Toyoaki Suzuki$^{2}$,\\
Bon-Chul Koo$^{3}$,
Ho-Gyu Lee$^{4}$,
Jae-Joon Lee$^{5}$,
and
Takashi Onaka$^{6}$
}
\institute{
Department of Physics,
Nagoya University, Furo-cho, Chikusa-ku, Nagoya, Aichi, 464-8602, Japan
\and
Institute of Space and Astronautical Science,
Japan Aerospace Exploration Agency,
3-1-1, Yoshinodai,
Sagamihara, Kanagawa, 252-5210, Japan
\and
Department of Physics and Astronomy,
Seoul National University,
599 Gwanak-ro, Gwanak-gu, Seoul 151-742, Korea
\and
Department of Astronomy and Astrophysics, University of Toronto,
Toronto, ON M5S 3H4, Canada
\and
Astronomy and Astrophysics Department, Pennsylvania State University,
University Park, PA 16802
\and
Department of Astronomy, Graduate School of Science,
University of Tokyo, 7-3-1 Hongo, Bunkyo-ku, Tokyo, 113-0033, Japan
}

\abstract
{}
{ 
We investigate
the spatial distribution of dust emission
around Tycho's SNR to understand its origin.
We distinguish
the dust associated with the SNR from
that of the surrounding ISM.
}
{ 
We performed mid- to far-infrared imaging observations
of the remnant at wavelengths of 9, 15, 18, 24, 65, 90, 140, and 160\,$\mu
$m
using the Infrared Camera and the Far-Infrared Surveyor onboard AKARI.
We compared the AKARI images with
the Suzaku X-ray image and the $^{12}$CO image of Tycho's SNR.
}
{ 
All the AKARI images except the 9, 140, and 160\,$\mu$m band images
show a shell-like emission structure
with brightness peaks at the
north east (NE) and 
north west (NW)
boundaries,
sharply outlining part of the X-ray shell.
The 140 and 160\,$\mu$m bands are dominated
by cold dust emission from the surrounding ISM near the NE boundary.
}
{ 
We conclude that
the dust emission at the NE boundary
comes from the ambient cloud interacting with the shock front,
while
the origin of the dust emission at the NW boundary
is rather unclear 
because of
the absence of prominent interstellar clouds
near the corresponding
region. We
cannot rule out the possibility that
the latter is mostly
of an SN ejecta origin.
}
\authorrunning{Ishihara, D. et al.}
\titlerunning{Dust emission from Tycho's SNR}

\bibpunct{(}{)}{;}{a}{}{,}

\begin{document}
\maketitle

\section{Introduction} \label{intro}

Tycho's supernova remnant (SNR; G120.1+01.4) is
the remnant of a type Ia supernova (SN) explosion 
\citep{Kraus},
which was observed by Tycho Brahe in 1572.
Its distance is estimated to be 1.5--3.1\,kpc
in several ways \citep{Kamper,Schuwaltz,Albinson,Strom}.
Tycho's SNR has been widely studied
in the X-ray and radio continua.
The remnant shows an 8$'$ diameter limb-brightened shell,
where the layer of the swept-up material
is very thin \citep{Warren}.
Inside the shell, metal line emissions in X-ray spectra
are produced predominantly by SN ejecta \citep{Bamba,XMM}.
\citet{Furuzawa} revealed the Fe-emitting ejecta
expanding at speeds of 2800--3350\,km\,s$^{-1}$ 
by the Suzaku observations.

In the synchrotron radio emission,
\citet{Reynoso} reported 
the current anisotropic expansion of the shock front
at speeds of 0\farcs45\,yr$^{-1}$ for the northwest (NW) 
and 0\farcs15\,yr$^{-1}$ for the northeast (NE) shock boundaries, 
based on the VLA observations over a 10\,yr interval.
\citet{Noveyama} suggested a possible interaction of this remnant
with a dense ambient cloud toward the NE direction
based on the $^{12}$CO observations.
\citet{Halpha} proposed the excitation of the pre-shock medium
around the NE boundary
by high-energy particles and/or fast neutral precursor 
from the H$\alpha$ observations.

In contrast,
infrared (IR) observations are relatively scarce,
although Tycho's SNR has been known as an IR emitter for a long time.
IRAS determined the IR flux of the remnant \citep{Saken},
while ISO resolved the dust emission
that originated in
collisional heating \citep{ISO} around the
shock front of Tycho's SNR.
The latter also indicate
that
a large amount of cold dust 
is not associated directly with the remnant.
In light of the chemical evolution of the universe,
type Ia SN plays an important role
in providing a significant fraction of the Fe group elements in the ISM.
However, the dust formation in type Ia SNR ejecta
has never been observed to date \citep[e.g.][]{Kepler},
while it has been reported for type II SNRs 
\citep[e.g.][]{ISO, CasA, HoGyu, BLAST} 
since the first detection from SN 1987A \citep[][]{Lucy,Mosley,Wooden}.

In this Letter, we present 
the latest fine and wide-area mid- to far-IR
AKARI images of Tycho's SNR.
By comparing them with X-ray and $^{12}$CO images,
we discuss 
the origin and physical state of the dust emission 
around the shell of Tycho's SNR.

\section{Observation and data} \label{obs}

We performed two pointed observations toward Tycho's SNR:
one with the Infrared Camera \citep[IRC;][]{IRC}
and the other with the Far-Infrared Surveyor \citep[FIS;][]{FIS}.
The AKARI mid-IR 15 and 24\,$\mu$m band imaging observations
were made on 2007 February 1 with the IRC.
The spatial resolution is $\sim$2\farcs5 for each image, 
where the pixel size is $\sim$2\farcs34$\times$2\farcs34.
The data were processed by using
the standard IRC imaging data reduction pipeline (version 20071017)$^1$.
The far-IR 65, 90, 140, and 160\,$\mu$m band images
were taken on 2007 February 1 with the FIS
in two round-trip slow scans \citep{FIS},
where the scan speed was 15$''$s$^{-1}$.
The spatial resolution is 30$''$ for 65 and 90\,$\mu$m
and 45$''$ for 140 and 160\,$\mu$m,
and the bin size of each image is set to be 25$''$.
The data were processed with the FIS Slow-Scan Toolkit (version 20070914)
\footnote{http://www.ir.isas.jaxa.jp/ASTRO-F/Observation/DataReduction/}
and corrected for cosmic-ray effects \citep{Jinken}.

The 9\,$\mu$m and 18\,$\mu$m band wide-area ($1^\circ\times1^\circ$)
images
around Tycho's SNR
were created from the mid-IR All-Sky Survey data \citep{ScanOpe}.
The original pixel scale of the image is 9\farcs36$\times$9\farcs36.
The data were processed 
by the pipeline developed for the point source catalog \citep{MirCat}, 
and additional custom procedures were applied in the same manner as
described in \citet{ic4954}.

The 0.4--10\,keV Suzaku/XIS X-ray image
was taken from the DARTS
\footnote{Data Archives and Transmission System (DARTS) at ISAS/JAXA
http://www.darts.isas.ac.jp/astro/suzaku/}
archives at ISAS/JAXA.
The $^{12}$CO(1$-$0) images were taken from the archives of
the Canadian Galactic Plane Survey \citep[CGPS;][]{CGPS}, where 
we integrated the data 
using the two velocity ranges: 
one is from $-$68\,km\,s$^{-1}$ to $-$55\,km\,s$^{-1}$
and the other is from $-$63\,km\,s$^{-1}$ to $-$60\,km\,s$^{-1}$. 
They are the likely maximal and minimal ranges
for the clouds associated and interacting with
the SNRs on the basis of the previous work \citep{Noveyama}.

\section{Results} \label{result}

\begin{figure*}
\includegraphics[width=18.5cm]{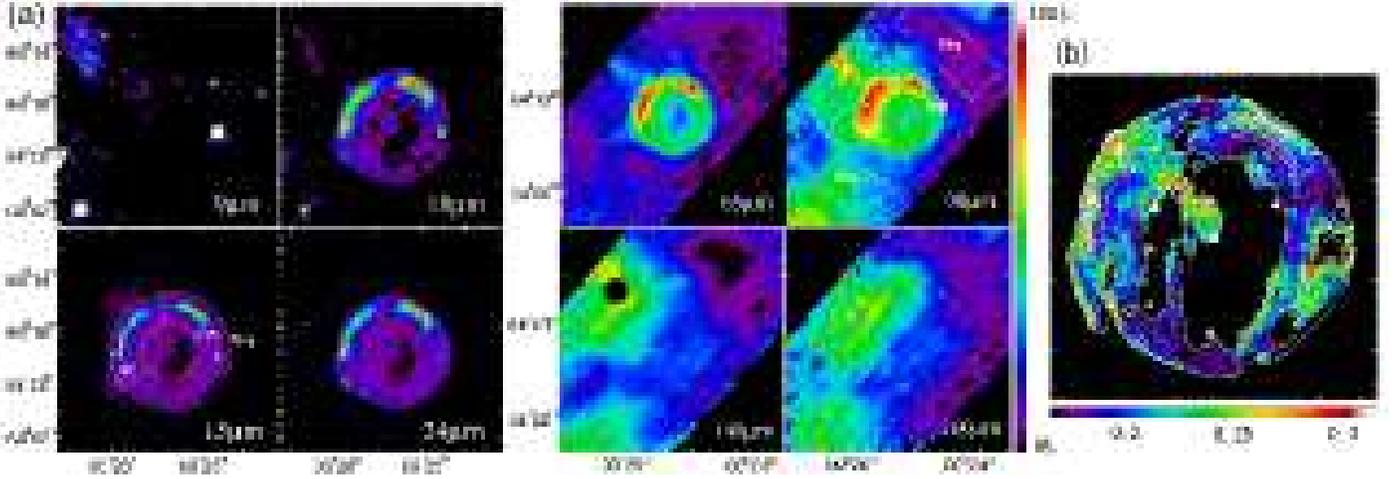}
\caption{(a) AKARI 4-band mid-IR and 4-band far-IR images
of Tycho's SNR drawn linearly 
in the color scales from zero to the maximum values of 5.3\,MJy sr$^{-1}
$, 27\,MJy sr$^{-1}$,
15\,MJy sr$^{-1}$, 62\,MJy sr$^{-1}$, 24\,MJy sr$^{-1}$, 26\,MJy sr$^{-1}$,
110\,MJy sr$^{-1}$, and 150\,MJy sr$^{-1}$
for the 9\,$\mu$m, 18\,$\mu$m, 15\,$\mu$m, 24\,$\mu$m, 65\,$\mu$m, 90\,
$\mu$m, 140\,$\mu$m, and 160\,$\mu$m band images,
respectively.
The 9\,$\mu$m and 18\,$\mu$m images are
created from the All-Sky Survey data,
making the spatial resolution and sensitivity
slightly worse than the 15\,$\mu$m and 24\,$\mu$m images 
from the pointed observation data.
The FOVs in the pointed observations were shown
in blue boxes
in the 15\,$\mu$m and 24\,$\mu$m images.
The definitions of the photometric apertures and sky regions
for 9--24\,$\mu$m bands and 65--160\,$\mu$m bands
in the 15\,$\mu$m and 90\,$\mu$m images,
respectively.
(b) The ratio map of the 15\,$\mu$m to the 24\,$\mu$m band.
The white contours indicate
the 4\%, 20\%, 34\%, 49\%, and 63\% levels of the peak brightness in
the 24\,$\mu$m emission.
The bright foreground star (GSC\,04023-01147) 
seen at 3\farcm5 away from the center to the west in the
panel (a)
is removed in the panel (b) and Figs.~\ref{fig:widearea} and \ref
{fig:warren}.
}\label{fig:akari}
\end{figure*}

Figure~\ref{fig:akari} shows
the AKARI multi-band images of Tycho's SNR, 
together with the ratio map of the 15\,$\mu$m to the 24\,$\mu$m band.
The 15\,$\mu$m, 18\,$\mu$m, and 24\,$\mu$m images
clearly show a limb-brightened shell-like structure
with several faint filaments.
There are strong emissions at around the NW and NE boundaries.
These bands contain important ionic line emissions
such as [Ne\,II] at 12.8\,$\mu$m and [Ne\,III] at 15.6\,$\mu$m,
and pure rotational lines
of molecular hydrogen
such as S(1)\,$J$=3$-$1 at 17.03\,$\mu$m
and S(2)\,$J$=4$-$2 at 12.28\,$\mu$m.
We estimated 
the contribution of the line emissions to the in-band fluxes
by convolving 
the Spitzer/IRS spectra of typical SNRs
\citep[W44, W28, 3C391, IC443;][]{Neufeld}
with the spectral response curves of the IRC \citep{IRC}.
As a result, the contribution of the total line emissions is
36--52\% for L15, 12--26\% for L18W, and 6--10\% for L24.
%
Therefore, there may be non-negligible contributions
from line emissions to the MIR intensities,
but a major fraction of the intensities come from the continuum emission.

The 9\,$\mu$m emission is relatively faint
and is not significantly detected
from the shell structure of the remnant.
The unusual faintness of the 9\,$\mu$m emission from the SNR
is clearly recognized
by making a comparison with the ISM cloud located at the NE 
corner in the 9\,$\mu$m and 18\,$\mu$m images. The
9\,$\mu$m/18\,$\mu$m 
brightness ratios are $<$0.003 at the boundary
and they are
$\sim$3 for the ISM cloud. 
The 9\,$\mu$m band includes
the polycyclic aromatic hydrocarbon (PAH) features 
at 6.2, 7.7, 8.6 and 11.3\,$\mu$m.
Their faintness is compatible with
the PAHs being
destroyed effectively in SNRs \citep[e.g.][]{Tielens}.
%

The 65\,$\mu$m and 90\,$\mu$m band images also show
the shell-like structure, 
while the 140\,$\mu$m and 160\,$\mu$m band images reveal
the dominance
of interstellar cold dust emission
around the NE boundary; 
the latter
does not appear
to be associated with the remnant.
The contribution of synchrotron emission to the far-IR fluxes is
negligible.
From the 20\,cm radio flux of 10--40\,mJy with a spectral index of $-$0.4
$\sim -$0.6 for each of the NE and NW regions \citep{Katz-stone}, we
estimate the contribution of synchrotron emission to be $<1\%$ of the
observed far-IR fluxes. 

\begin{table}
\caption{Infrared flux and its properties 
of the NW and NE regions of Tycho's SNR.}
\label{tbl:sed}
\begin{center}
\begin{tabular}{cccc}\hline\hline
Wavelength$^*$&\multicolumn{3}{c}{Flux (Jy)}\\
($\mu$m)    & Total         & NW            & NE             \\\hline
8.61 (4.10) & $<$0.05       & $<$0.05       & $<$0.05 \\
15.6 (5.98) & 5.9$\pm$4.8   & 2.0$\pm$1.1   &  2.5$\pm$1.2 \\
18.4 (9.97) & 14.8$\pm$4.4  &  5.1$\pm$1.5  &  4.2$\pm$1.3 \\
 22.9 (5.34)&28.3$\pm$3.2   & 9.5$\pm$0.8   & 10.7$\pm$0.8 \\
 65  (21.7) &38.6$\pm$7.7   & 5.7$\pm$1.1   & 14.4$\pm$2.9 \\
 90  (37.9) &40.6$\pm$8.1   & 8.5$\pm$1.7   & 16.8$\pm$3.4 \\
140  (52.4) &46.7$\pm$14.0  & 5.1$\pm$1.5   & 30.0$\pm$9.0 \\
160  (34.1) &44.5$\pm$13.4  & 4.6$\pm$1.4   & 38$\pm$12 \\
\hline
\multicolumn{4}{c}{Fitting results}\\
\hline
T$_1(K)$  & 95$\pm$2 & 109$\pm$7& 107$\pm$4 \\
T$_2(K)$  & 25$\pm$1 & 25$\pm$10 & 22$\pm$1 \\
$M_{\rm warm\ dust}$($\times$10$^{-4}$${\rm M}_\odot$)
& 10$\pm$3 
& 2.0$\pm$0.4
& 2.0$\pm$0.6\\
$M_{\rm cold\ dust}({\rm M}_\odot)$
&0.3$\pm$0.1 
&0.03$\pm$0.01
&0.3$\pm$0.1\\
\hline
\end{tabular}\\
\end{center}
$*$ The numbers in the parentheses are effective bandwidths
\citep{IRC,FIS}.\\
\end{table}

\begin{figure}
\center
\includegraphics[width=7cm]{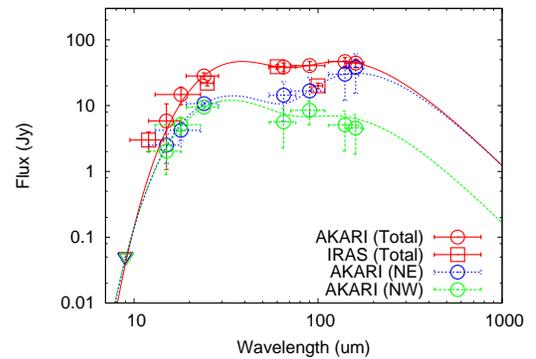} 
\caption{SEDs for the total (red solid line),
the NE (blue dashed), 
and the NW region (green dotted) of Tycho's SNR, 
fitted by a two-temperature graybody model.
The AKARI and IRAS measurements are indicated 
by the open circles and the open squares, respectively.
}
\label{fig:sed0}
\end{figure}

We derived the flux densities of Tycho's SNR for the total,
the NE, and the NW 
regions,
separately, and
definitions of the photometric apertures are shown
in the 15\,$\mu$m and 90\,$\mu$m images in Fig.~\ref{fig:akari}a.
For each region, 
the
resulting
spectral energy distribution (SED)
is then fitted by a two-temperature graybody model (Fig.~\ref{fig:sed0}).
The temperatures and masses of the dust thus derived,
as well as the flux densities,
are summarized in Table~\ref{tbl:sed}.
In the estimate of the dust mass,
we assume a dust mass absorption coefficient of 28\,cm$^2$\,g$^{-1}$ at 90\,$\mu$m 
\citep{Hildebrand}.
%
%

As seen in Fig.~\ref{fig:sed0}, the AKARI measurements
for the SED of the entire remnant show
overall agreement with the previous ones \citep{Saken}
except at 100\,$\mu$m,
where IRAS gives a significantly
lower value.
We suspect that
the IRAS flux is affected
by the presence of the cold dust emission
around the NE boundary
in subtracting sky background.
All the SEDs are reproduced 
with dust temperatures of $T_1\sim100$\,K and $T_2\sim$20\,K, and
the former can be interpreted
by collisionally heated dust in the postshock plasma.
The SED of the NE region, however, needs 
a large amount (0.3\,M$_\odot$)
of cold ($\sim$20\,K) dust, which is likely attributed to 
the pre-existing ISM
as already suggested by \citet{ISO}
and spatially resolved in AKARI far-IR images
(Fig.~\ref{fig:akari}a).

Figure~\ref{fig:akari}b shows
the ratio map of the 15\,$\mu$m to the 24\,$\mu$m band.
The ratios were calculated
after
subtracting
background in each band,
which was estimated by averaging the brightness of 
blank-sky areas surrounding the SNR.
In this map, the ratios of 0.18 and 0.46 correspond to dust temperatures
of 100\,K and 136\,K, respectively, 
when we assume that these bands are dominated by 
thermal dust emission with the emissivity power-law index of 1.0.
Thus these values are roughly consistent with the above warmer dust
component.
The figure shows that
the dust temperature reaches a local maximum near the shock front,
decreasing toward inner regions.
As a whole, 
high-temperature regions are distributed more isotropically
around the shell
than the dust emission itself,
suggesting that
the dust is shock-heated at the shell boundary.
The systematic decrease
in the ratio toward inner regions may indicate that
smaller grains are mainly
destroyed by sputtering.

\section{Discussion} \label{discussion}

\begin{figure}
\center
\includegraphics[width=8.5cm]{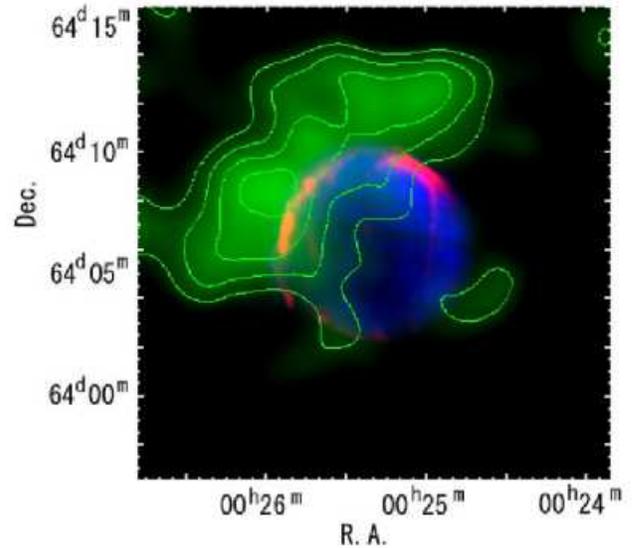}
\caption{Composite image of Tycho's SNR, consisting of
AKARI 24\,$\mu$m band intensity in red,
the CGPS $^{12}$CO intensity integrated over the velocity range
from $-$68 to $-$55\,km s$^{-1}$ in green, 
and Suzaku 0.4--10\,keV X-ray intensity in blue,
The X-ray intensity is in the range of
2.6--13$\times$10$^{-5}$ counts s$^{-1}$,
and the 24\,$\mu$m intensity in the range of 4--40 MJy sr$^{-1}$.
The contours correspond to the brightness temperatures of
0.28, 0.43, 0.57, and 0.71\,K
for the CO integrated intensity.}
\label{fig:widearea}
\end{figure}
%
%
Figure~\ref{fig:widearea} shows
the composite images of Tycho's SNR,
consisting of the AKARI mid-IR (24\,$\mu$m),
the Suzaku X-ray (0.4--10\,keV), and the CGPS $^{12}$CO images.
In the figure, the X-ray emission, which shows
a circular outer shell,
is surrounded by the $^{12}$CO emission.
They depict the past isotropic expansion and
the current interaction with the dense medium at the NE boundary.
Part of the X-ray shell is sharply outlined
by the dust emission.
We discuss the origins of these dust emissions below.
First, we show that 
the dust 
cannot be totally destroyed by sputtering
in the post-shock hot plasma.
The dust have resided in the plasma for $\sim$50\,yr
with the current shock speed and the thickness of the IR emission region,
while the sputtering destruction timescale of the grains
with a radius as
small as 1\,nm is
estimated to be 150\,yr
\citep{tau} for 
the plasma temperature and density of $8.4\times10^{6}$\,K
and 10\,cm$^{-3}$ \citep{Warren}, respectively.

As described above,
strong cold dust emission toward the NE direction
is detected in the 140 and 160\,$\mu$m images
(Figs.~\ref{fig:akari}a).
Indeed, the SED at the NE boundary
exhibits the presence of a large amount of cold dust (Fig.~\ref{fig:sed0}).
The distribution of the cold dust shows
a spatial correspondence with the $^{12}$CO(1$-$0) cloud
surrounding the NE part of the remnant (Fig.~\ref{fig:widearea}),
which are probably located closely to the SNR
from the $^{12}$CO line velocities. 
The mid-IR dust emission comes from 
the outer edge of the cold dust and molecular cloud.
Thus for the NE boundary, 
the mid-IR emission is very likely to
originated in
the shock-heated dust
through interaction of the SNR with the ambient cloud.
Unlike the NE boundary, 
the NW boundary region does not show
the clear presence of ISM clouds
in the cold dust and $^{12}$CO emissions.
There are no HI clouds
around the corresponding region \citep{Reynoso9}.
One possibility is that clouds,
which were present there, have already been
dispersed into ionized gas,
and only dust remains.
%
%
The highly anisotropic morphology,
however, may not favor this scenario.

To make a more quantitative comparison 
between the NE and the NW boundary,
we estimate the molecular gas masses
using the same apertures as in Fig.~\ref{fig:akari}a 
with the $^{12}$CO X-factor of a typical Galactic value, 
3$\times$10$^{20}$\,cm$^{-2}$ K$^{-1}$ km$^{-1}$ s.
Integrating over the velocity range
of $-$68 to $-$55\,km s$^{-1}$,
we derive the gas masses of 90 and 30\,M$_\odot$
in the NE and the NW region, respectively, 
while they are 20 and 2\,M$_\odot$
for the velocity range of $-$63 to $-$60\,km s$^{-1}$.
Thus, from Table~\ref{tbl:sed},
the gas-to-dust mass ratios are $70-300$ in NE 
and $70-1000$ in NW for the cold dust,
while they are $(1-5)\times 10^5$ in NE and 
$(0.1-2)\times 10^5$ in NW for the warm dust. 
As for the cold dust, the ratios
in NE and NW are similar to each other
around a typical ISM value 
\citep[100--200;][]{cirrus}, 
although the NW ratio has a large uncertainty 
depending on the adopted velocity range.
Therefore the cold dust is likely to be
of a pre-existing ISM origin.
However for the warm dust,
which is collisionally heated by the SNR, 
the gas-to-dust mass ratio in NW is
systematically less
than in the NE.
If we adopt
more restricted
velocity range,
which is more appropriate 
to the interacting part of the cloud,
the difference is
as much as a factor of 10. 
Therefore we conclude that
the NW region is relatively rich in warm dust.

%
%
%
%
%

\begin{figure*}
\center
\includegraphics[width=13cm]{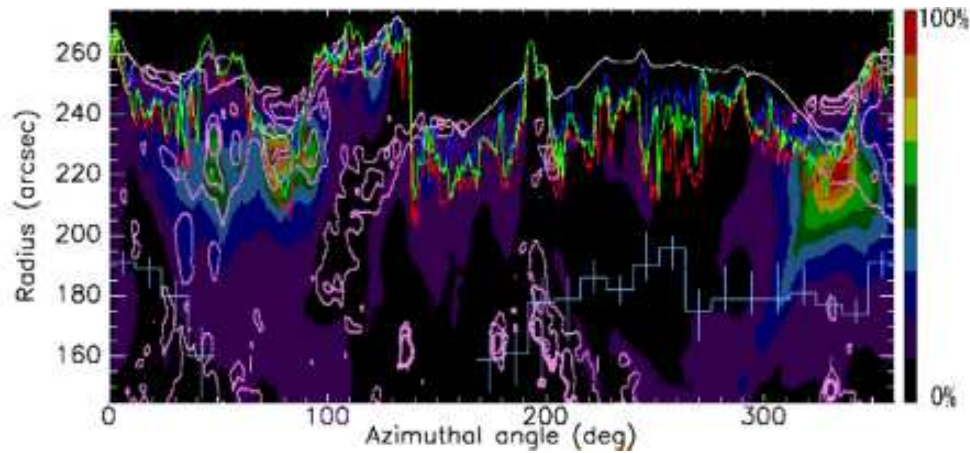}
\caption{AKARI 24\,$\mu$m intensity in the color, 
as well as the magenta contours
of the 15\,$\mu$m to 24\,$\mu$m band ratio,
are shown in a plane of the radius versus azimuthal angle. 
The color scale is linear with a maximum value of 62\,MJy sr$^{-1}$,
while the contour levels correspond to the band ratios
of 0.24, 0.26, and 0.28.
The center position is 
R.A.= 00$^{\rm h}$25$^{\rm m}$19$^{\rm s}$.40,
decl.=+64$^\circ$08$'$13\farcs98.
The azimuthal angle is measured from the north through the east.
For comparison,
the positions of the blast wave (the top white curve),
the contact discontinuity
(2$\sigma$: red, 3$\sigma$:green, 4$\sigma$:blue), 
and the reverse shock (cyan) are shown together,
which are taken from \citet{Warren}.
}
\label{fig:warren}
\end{figure*}

Figure~\ref{fig:warren} shows
the distribution of the dust emission
compared to the positions 
of the blast wave,
the contact discontinuity (CD),
and the reverse shock,
all of which are plotted
in a plane of the radius versus azimuthal angle. 
The NE bright spot at the angle of $\sim$80$^\circ$ is located
between the forward shock and the CD,
which is compatible with the picture of
the NE dust mostly having
an interstellar origin (i.e. swept-up materials).
A local maximum of the dust temperature coincides
well with the brightness peak.
In contrast,
the NW bright spot ($\sim$330$^\circ$) is located
between the reverse shock and the CD.
The dust temperature does not peak
at the NW bright spot (see also Fig.~\ref{fig:akari}b).
Thus the situation is considerably different
between
the NE and NW bright spots.
Judging from these,
combined with the filamentary structures
extended toward inner regions,
a majority of
the NW dust could have
an SN ejecta origin.

It should be noted that
the projection effect can explain the difference 
because the swept-up layer of this remnant is very
thin; however
\citet{Warren} showed Si-rich X-ray spectra
from the NW region where the mid-IR dust emission is bright,
suggesting that the emitting matter
comes from ejecta
rather than from ISM.
With AKARI,
\citet{HoGyu} detected dust emission
from ejecta of the type II SNR, G292.0+1.8,
which has the 15\,$\mu$m to 24\,$\mu$m band ratio of $>$0.5.
If the NW dust emission also comes from ejecta,
the band ratio is significantly lower
than the above ratio,
which may possibly be attributed
to the difference in chemical composition
of dust between Types Ia and II SNRs.

\section{Summary}
We have presented the latest fine and wide-area mid- to far-IR
AKARI images of Tycho's SNR, which are compared with the 
X-ray and $^{12}$CO images.
They show a shell-like emission structure
with bright peaks at the NE and NW boundaries,
sharply outlining part of the X-ray image.
Most of the mid-IR dust emission comes from the dust shock-heated 
at the shock front. A significant fraction 
of the far-IR and the PAH emission comes from the ISM clouds 
near the NE shock boundary, which further reveals a large-scale 
jet-like structure in front of the NE boundary with a spatial 
correspondence in the $^{12}$CO emission. 
We conclude
that the mid-IR dust emission at the NE boundary
comes from the ISM interacting with the shock front,
judging from the spatial correlation
among the mid-IR dust, the cold dust, and the molecular cloud. 
The origin of the dust emission at the NW boundary is rather unclear
due to the absence of prominent interstellar clouds
near the corresponding region.
We
estimated gas-to-dust mass ratios at the NE and the NW 
boundary to reveal that the NW region is
very
rich in warm dust.
We find that a large fraction of the NW dust emission comes from the region 
between the reverse shock and the CD,
assuming there is no projection effect.
We
therefore cannot
rule out the possibility that
a major fraction of
the dust emission at the NW boundary is of
an SN ejecta origin.

\begin{acknowledgements}
This research is based on observations with AKARI,
a JAXA project with the participation of ESA.
This study was initiated by preliminary studies done by high school
students 
who visited Nagoya University for a one-week internship. 
We have made use of the NASA/IPAC Infrared Science Archive,
which is operated by the Jet Propulsion Laboratory,
California Institute of Technology,
under contract with the National Aeronautics and Space Administration,
and archival data from the Canadian Galactic Plane Survey (CGPS),
a Canadian project with international partners,
supported by the Natural Sciences and Engineering Research Council.
We also thank A. Kawamura for providing
precious suggestions
for analyzing
of the molecular clouds.
This work was supported by
the the Nagoya University Global COE Program, 
``Quest for Fundamental Principles in the Universe (QFPU)''
from JSPS and MEXT of Japan.
We also express many thanks to the 
anonymous referee for
a careful reading
and constructive comments.

\end{acknowledgements}


\begin{thebibliography}{Mittelbach100}
\bibitem[Albinson et al.(1986)]{Albinson}
Albinson, J. S., Tuffs, R. J., Swinbank, E., Gull, S. F. 1986, \mnras, 219,
427
\bibitem[Bamba et al.(2005)]{Bamba}
Bamba A., Yamazaki, R., Yoshida, T., 2005, \apj, 621, 793
\bibitem[Blair et al.(2007)]{Kepler} 
Blair, W. P., Ghavamian, P., Long, K.S., et al., 2007, \apj, 662, 998
\bibitem[Decourchelle et al.(2001)]{XMM}
Decourchelle, A., Sauvageot, J.L., Audard, M., et al., 2001, \aap, 365,
L218
\bibitem[Douvion et al.(2001)]{ISO}
Douvion, T., Lagage, P.O., Cesarsky, C.J., Dwek, E., 2001, \aap, 373, 281
\bibitem[Furuzawa et al.(2009)]{Furuzawa}
Furuzawa, A., Ueno, D., Hayato, A., et al., 2009, \apj, 693, L61
\bibitem[Ghavamian et al.(2000)]{Halpha}
Ghavamian, P., Raymond, J., Hartigan, P., Blair, W. 2000, \apj, 535, 266
\bibitem[Ishihara et al.(2006)]{ScanOpe}
Ishihara, D., Wada, T., Onaka, T., et al., 2006a, \pasp, 118, 324
\bibitem[Ishihara et al.(2007)]{ic4954}
Ishihara, D., Onaka, T., Kaneda, H., et al., 2007, \pasj, 59, S443
\bibitem[Ishihara et al.(2010)]{MirCat}
Ishihara, D., Onaka, T., Kataza, H., et al., 2010, \aap, 514, 1
\bibitem[Hildebrand(1983)]{Hildebrand}
Hildebrand, R. H., 1983, \qjras, 24, 267
\bibitem[Kamper \& van den Bergh(1978)]{Kamper}
Kamper, K. W., and van den Bergh, S., 1978, \apj, 224, 851
\bibitem[Kawada et al.(2007)]{FIS}
Kawada, M., Baba, H., Barthel, P. D., et al. 2007, \pasj, 59, S389
\bibitem[Katz-Stone et al.(2000)]{Katz-stone}
Katz-Stone, D. M., Kassim, N. E., Lazio, T. J. W., O'Donnell, R. 2000,
\apj, 529, 453
\bibitem[Krause et al.(2008)]{Kraus}
Krause, O., Tanaka, M., Usuda, T., et al., 2008, \nat, 456, 617
\bibitem[Lee et al.(2009)]{HoGyu} 
Lee, H.G., Koo, B.C., Moon, D.S., et al., 2009, \apj, 706, 441
\bibitem[Lee et al.(2004)]{Noveyama}
Lee, J.J., Koo, B.C., Tatematsu, K. 2004, \apj, 605, L113
\bibitem[Lucy et al.(1989)]{Lucy}
Lucy, L. B., Danziger, I. J., Gouiffes, C., Bouchet, P. 1989, LNP, 350,
164
\bibitem[Mosley et al.(1989)]{Mosley}
Moseley, S. H., Dwek, E., Glaccum, W., Graham, J. R., Loewenstein, R. F.,
1989, \nat, 340, 697
\bibitem[Neufeld et al.(2007)]{Neufeld}
Neufeld D. A., Hollenbach D. J., Kaufman, M.J., et al., 2007, \apj, 664,
890
\bibitem[Onaka et al.(2007)]{IRC}
Onaka, T., Matsuhara, H., Wada, T., et al. 2007, \pasj, 59, S401
\bibitem[Reynoso et al.(1997)]{Reynoso}
Reynoso, E. M., Moffett, D. A., Goss, W. M., et al., 1997, \apj, 491, 816
\bibitem[Reynoso et al.(1999)]{Reynoso9}
Reynoso, E. M., Vel\'{a}zquez, P. F., Dubner, G. M., and Goss, W. M.
1999, \apj, 117, 1827
\bibitem[Rho et al.(2008)]{CasA} 
Rho, J., Kozasa, T., Reach, W.R., et al., 2008, \apj, 673, 271
\bibitem[Saken et al.(1992)]{Saken}
Saken, J. M., Fesen, R. A., Shull, J. M. 1992, \apjs, 81, 715
\bibitem[Schwarz et al.(1995)]{Schuwaltz}
Schwarz, U. J., Goss, W. M., Kalberla, P. M., Benaglia, P. 1995, \aap, 299,
193
\bibitem[Sibthorpe et al.(2010)]{BLAST}
Sibthorpe, B., Ade, P. A. R., Bock, J. J., et al. 2010, \apj, in press
\bibitem[Sodroski et al.(1994)]{cirrus}
Sodroski, T. J., Bennett, C., Boggess, N., et al., 1994, \apj, 428, 638
\bibitem[Strom(1988)]{Strom}
Strom, R. G., 1988, \mnras, 230, 331
\bibitem[Suzuki et al.(2007)]{Jinken}
Suzuki, T., Kaneda, H., Nakagawa, T., et al., 2007, \pasj, 59, 473
\bibitem[Tielens et al.(1994)]{tau}
Tielens, A. G. G. M., McKee, C. F., Seab, C. G., Hollenbach, D. J. 
1994, \apj, 431, 321
\bibitem[Taylor et al.(2003)]{CGPS}
Taylor, A. R., Gibson, S. J., Peracaula, M., et al., 2003, \aj, 125, 3145
\bibitem[Tielens et al.(2008)]{Tielens}
Tielens, A. G. G. M., 2008, \araa, 46, 289
\bibitem[Warren et al.(2005)]{Warren}
Warren, J. S., Hughes, J. P., Badenes, C., et al., 2005, \apj, 634, 376
%
240, 7
%
\bibitem[Wooden(1993)]{Wooden}
Wooden 1993, \apjs, 88, 477
\end{thebibliography}
\end{document}